\documentclass[prb,twocolumn,showpacs]{revtex4-1}
\usepackage{epsfig}
\usepackage{graphicx}
\usepackage{amsfonts,amssymb}
\usepackage{amsmath}
\usepackage{color}
\usepackage{times}
\definecolor{refcolor}{rgb}{1.0,0.0,0.0}

\newcommand{\be}{\begin{equation}}
\newcommand{\ee}{\end{equation}}   
\newcommand{\bea}{\begin{eqnarray}}
\newcommand{\eea}{\end{eqnarray}}
\newcommand{\ba}{\begin{array}}
\newcommand{\ea}{\end{array}}

\newcommand{\q}{{\bf q}}
\renewcommand{\k}{{\bf k}}
\newcommand{\Q}{{\bf Q}}

\begin{document}

\title{Spin-wave excitations in the SDW state of iron pnictides: a comparison between the roles of interaction parameters}
\author{Dheeraj Kumar Singh}
\email{dheerajsingh@hri.res.in} 
 
\affiliation{Harish-Chandra Research Institute, Chhatnag Road, Jhunsi, Allahabad 211019, India }
\affiliation{Homi Bhabha National Institute, Training School Complex, Anushakti Nagar, Mumbai 400085, India}
\begin{abstract}
We investigate the role of Hund's coupling in the spin-wave excitations of the ($\pi, 0$) ordered magnetic state
within a five-orbital tight-binding model for iron pnictides. To differentiate between the roles of intraorbital 
Coulomb interaction and Hund's coupling, we focus on the self-consistently obtained mean-field SDW state with 
a fixed magnetic moment obtained by using different sets of interaction parameters. We find that the Hund's coupling 
is crucial for the description of various experimentally observed characteristics of the spin-wave
excitations including the anisotropy, energy-dependent behavior, and spin-wave spectral weight distribution.
\end{abstract}
\pacs{74.70.Xa,75.30.Ds,75.30.Fv}
\maketitle
\newpage
\section{Introduction}
Iron-based superconductors are prototypes of moderately correlated\cite{qazilbash,yang} and multiorbital
systems exhibiting unconventional superconductivity(SC).\cite{stewart} A close
proximity of the superconducting phase to the antiferromagnetic (AFM) phase\cite{avci, hussey} indicates
a crucial role of the spin fluctuations like in the cuprate superconductors. The pairing mediated by the 
spin-fluctuations has been a subject of intense investigation both theoretically and 
experimentally.\cite{scalapino,dahm} An experimental signature of the role of spin-fluctuations, for instance, can
be obtained in the form of change in magnetic-exchange energy involved 
in the transition from the normal to the superconducting state, which can then be compared with 
the superconducting condensation energy. The magnetic-exchange energy, on the other hand, can be 
estimated from the measurements of the spin excitations that yields information 
regarding the exchange coupling present in the system.\cite{zhou,wang} Therefore, it is of much importance to comprehend 
various characteristics of the spin excitations for gaining insight into the mechanism
of the Cooper pair formation.

Parent compounds of iron pnictides exhibit a spin-density wave state
with ordering wavevector $\Q = (\pi, 0)$, which involves parallel ferromagnetic chains running along 
$y$ direction while being coupled to each other antiferromagnetically along $x$.\cite{zhao} This
particular spin arrangement arises due to the significant nesting present between the circular 
hole pockets at $\Gamma$ and the elliptical electron pockets at X in the Fermi surfaces (FSs) of 
the unfolded Brillouin zone corresponding to one Fe atom per unit
cell.\cite{mazin,singh,haule,yi,kondo,yi1,brouet,kordyuk} Existence of the Fermi surfaces in
the SDW state,\cite{yi,yi1} metallicity\cite{chuang,nakajima} and 
small magnetic moments\cite{cruz,huang} with largest being $\approx$ 1$\mu_B$ found in 122 series supports the 
nesting based scenario. However, experimental investigation carried out by the inelastic neutron scattering
(INS) measurements reveals remarkably high-energy scale of the spin-wave excitations, which are 
sharp, highly dispersive, and can extend up to energy $\sim$ 200 meV.\cite{ewings,diallo,zhao1,harriger,ewings1}

The spin-wave dispersion can be  
described within conventional Heisenberg model with highly anisotropic exchange couplings.\cite{yao} However, such a 
description suffers from the limitation that it 
cannot explain the spin-wave damping resulting from 
the particle-hole excitations in the metallic SDW state. The limitation may be
overcome by considering additional terms which 
account for the bandstructure and coupling between the electron spin and the local spin through 
Hund's coupling.\cite{lv} Various studies give an estimate of the intraorbital Coulomb interaction ($U$) to be 
$U$/$W \approx$ 0.25, where $W$ is the bandwidth.\cite{yang,anisimov} Therefore, plausibly a completely
itinerant approach is best suited to describe the spin excitations in
these materials. Within the latter approach, excitonic\cite{brydon} and orbital\cite{nimisha,knolle} models 
have been frequently employed to 
investigate the spin-wave excitations. A comparative analysis of these two models while considering a two-orbital model 
has also been carried out, where excitations were found to be heavily damped away from the ordering wavevector $\Q$ in
the latter.\cite{knolle} 

Standard on site interaction includes two important parameters - intraorbital Coulomb 
interaction $U$ and Hund's coupling $J$. The correlation effect due to $U$ involves the suppression of charge 
fluctuations whereas that due to $J$ pertains to the unscreened high-spin state on the neighboring sites
resulting in the correlations. Different values of $U/J$ have often been used by different groups based on 
the different estimates from different methods. For instance, a combination of the constrained
random-phase approximation and the maximally localized
Wannier function yields $J/U \sim$ 0.14 \cite{miyake} whereas work based on the dynamical mean-field
theory estimates is $J/U \sim 0.25$.\cite{ishida} Similarly, experiments also provide varying estimates for
the same.\cite{schafgans,yang} With regard to various properties of this material, a comparative role
of these two interaction parameters is of strong current interest.

$U$ vs $J$ phase diagrams in this direction is an important step.\cite{luo} Furthermore, in the two-orbital model, it has been illustrated that $J$ plays an important role in
stabilizing the doped SDW state against long-wavelength fluctuations through the generation of 
additional ferromagnetic spin coupling involving the inter-orbital susceptibility.\cite{nimisha} 
The spin excitations have been studied recently within the five-orbital model 
as function of $U$ but with a fixed Hund's coupling $J$.\cite{kaneshita,kovacic} Well-defined branches extending
upto high energy were obtained for those values of $U$ that led to the magnetic moment $m \sim 1$ or larger. An 
important issue that has not attracted much attention is the role of $J$ in the various features of 
spin-wave excitations such as dispersion, anisotropy and the way spin-wave spectral weight is 
distributed.

In this paper, we examine the role of Hund's coupling in the spin-wave excitations of ($\pi, 0$) SDW state of undoped 
iron pnictides within a five-orbital tight-binding model. The interaction parameters are 
chosen in such a way that a fixed magnetic moment $m \approx 1$ results in the self-consistent SDW state, which 
is motivated by the observed magnetic moments in 122 compounds. we find that $J$ 
is crucial for (i) the sharp and well-defined excitations
up to high energy, (ii) the anisotropy in the excitations around X, and
(iii) the fact that the spin-wave spectral weight is concentrated near $\sim$ 200meV.

The plan of the paper is as follows. In section II, a mean-field description of the ($\pi, 0$) SDW as well as the 
strategy to calculate spin-wave excitations is presented. In section III, results on the spin-wave dispersion
along high-symmetry direction for
different combination of intraorbital Coulomb interaction and Hund's coupling are presented. The way spin-wave spectral 
function behaves as a function of the interaction parameters is also discussed. Finally, conclusions are presented in 
section IV.
\section{Theory}
The kinetic part of the model Hamiltonian that we consider is 
\begin{equation}
 \mathcal{H}_0 = \sum_{\k}\sum_{\mu,\nu}\sum_{\sigma} \varepsilon_{\k}^{\mu\nu} 
d_{\k \mu\sigma}^\dagger d_{\k \nu\sigma} + \text{H.c.}, 
\end{equation} 
where the operator $d_{{\bf k} \mu \sigma}^\dagger$ ($d_{{\bf k} \mu \sigma}$) creates (destroys)
an electron in the $\mu$-th orbital with spin $\sigma$ and momentum $\k$, and
$\varepsilon_{\k}^{\mu\nu}$ are the hopping elements\cite{ikeda} from orbital $\mu$ to $\nu$. The orbitals $\mu$ 
and $\nu$ belong to the set of five $d$-orbitals $d_{xz}$, $d_{yz}$, $d_{xy}$, $d_{x^2-y^2}$, and $d_{3z^2-r^2}$.

Standard onsite Coulomb interactions in the Hamiltonian
\begin{eqnarray}
\mathcal{H}_{int} &=& U \sum_{{\bf i},\mu} n_{{\bf i}\mu \uparrow} n_{{\bf i}\mu \downarrow} + (U' -
\frac{J}{2}) \sum_{{\bf i}, \mu<\nu} n_{{\bf i} \mu} n_{{\bf i} \nu} \nonumber \\ 
&-& 2 J \sum_{{\bf i}, \mu<\nu} {\bf{S_{{\bf i} \mu}}} \cdot {\bf{S_{{\bf i} \nu}}} + J \sum_{{\bf i}, \mu<\nu, \sigma} 
d_{{\bf i} \mu \sigma}^{\dagger}d_{{\bf i} \mu \bar{\sigma}}^{\dagger}d_{{\bf i} \nu \bar{\sigma}}
d_{{\bf i} \nu \sigma} \nonumber\\
\label{int}
\end{eqnarray}
include the intraorbital (interorbital) Coulomb interaction term as the first (second) term. 
The last two terms represent the Hund’s coupling and the pair
hopping energy, respectively.  

The Hamiltonian in the ($\pi, 0$) SDW state after mean-field approximation is obtained as
\be
{\mathcal{H}}_{\k} = 
\sum_{\k \sigma}\Psi^{\dagger}_{{\bf k} \sigma}
\begin{pmatrix}
 \hat{\varepsilon}_{\k}+\hat{N} \,& \,{\rm sgn}\bar{\sigma}\hat{\Delta} \\
 {\rm sgn}\bar{\sigma}\hat{\Delta} \,& \,\hat{\varepsilon}_{\bf {k+Q}}+\hat{N}
\end{pmatrix} 
\Psi_{{\bf k} \sigma},
\ee
where $\Psi^{\dagger}_{{\bf k} \sigma} = (c^{\dagger}_{{\k}1\sigma},....,c^{\dagger}_{{\k}5\sigma},
{c}^{\dagger}_{{\k}\bar{1}\sigma},....,{c}^{\dagger}_{\k\bar{5}\sigma})$ with $\bar{c}^{\dagger}_{{\k}\bar{\mu}\sigma}$ = 
$c^{\dagger}_{{\k+\Q}\mu\sigma}$. Matrix elements of matrices $\hat{M}$ and $\hat{N}$ are 
\bea
2\Delta_{\mu\mu} &=& Um_{\mu\mu}+J\sum_{\mu \ne \nu}m_{\nu\nu} \nonumber\\
2\Delta_{\mu\nu} &=& Jm_{\mu\nu}+(U-2J)m_{\nu\mu}
\eea
and 
\bea
2N_{\mu\mu} &=& Un_{\mu\mu}+(2U-5J)\sum_{\mu \ne \nu}n_{\nu\nu} \nonumber\\
2N_{\mu\nu} &=& Jn_{\mu\nu}+(4J-U)n_{\nu\mu},
\eea
where 
\be
n_{\mu\nu} = \sum_{\k \sigma} \langle c^{\dagger}_{\k \mu \sigma}c_{\k \nu \sigma}\rangle,  \,\,\, 
m_{\mu\nu} = \sum_{\k \sigma} \langle c^{\dagger}_{\k \bar{\mu} \sigma}c_{\k \nu \sigma}\rangle .
\ee

Multiorbital transverse-spin susceptibility is defined as 
\bea
&&\chi_{\alpha \beta, \mu \nu}(\q,\q^{\prime},i\omega_n)  \nonumber\\
&&=\frac{1}{\beta} \int^{\beta}_0{d\tau e^{i \omega_{n}\tau}\langle T_\tau
[{ S}^{+}_{\alpha \beta}(\q, \tau) {S}^{-}_{\nu \mu} (-\q^{\prime}, 0)]\rangle}.
\eea
Thus,
\bea 
&&\chi^{}_{\alpha \beta, \mu \nu}(\q,\q,i\omega_n)  \nonumber\\
&&= \sum_{\k,i\omega^{\prime}_n} G^{0\uparrow}_{\alpha \mu}
(\k+\q,i\omega^{\prime}_n+i\omega_n)G^{0\downarrow}_{\nu \beta} (\k,i\omega^{\prime}_n).
\eea
The components of the spin operator in Eq. 7 are given by
\be
{\cal S}^{i}_{\q}= \sum_{\bf k} \sum_{\sigma \sigma^{\prime}} \sum_{\mu \mu^{\prime}} d^{\dagger}_{\mu \sigma}(\k+\q)
E_{\mu \mu^{\prime}} \sigma^{i}_{\sigma \sigma^{\prime}} d_{\mu^{\prime} \sigma^{\prime}}(\k),
\ee
where $i = x,y,z$, $\hat{E}$ is a 5$\times$5 identity matrix corresponding to the orbital bases, and $\sigma^i$s are the Pauli matrices for 
the spin degree of freedom. 

An element of the transverse-spin susceptibility 
in the SDW state is
\bea
{\chi}^{0}_{\alpha \beta, \mu \nu} = \chi^{}_{\alpha \beta, \mu \nu}
+\chi^{}_{\bar{\alpha} \beta, \bar{\mu} \nu}+\chi^{}_{\alpha \bar{\beta}, \mu \bar{\nu}}
+\chi^{}_{\bar{\alpha} \bar{\beta}, \bar{\mu} \bar{\nu}},
\eea
Then, the susceptibility matrix can be written as 
\be
\hat{{\chi}}^0(\q,i\omega_n) = 
\begin{pmatrix}
 \hat{{\chi}}^0(\q,\q,i\omega_n) \,& \,\hat{{\chi}}^0(\q,\q+\Q,i\omega_n) \\
 \hat{{\chi}}^0(\q+\Q,\q,i\omega_n) \,& \,\hat{{\chi}}^0(\q+\Q,\q+\Q,i\omega_n)
\end{pmatrix}, 
\ee
where $\hat{{\chi}}^0(\q,\q,i\omega_n)$ and others are $n^2\times n^2$ matrices. 
Physical transverse-spin susceptibility corresponding to the spin operators defined by Eq. 8 is  
\be
{\chi}^{ps}(\q, i\omega_n) = \sum_{\alpha \mu} {\chi}^0_{\alpha \alpha, \mu \mu} (\q, \q, i\omega_n).
\ee

Interaction effects are incorporated within the random-phase approximation (RPA) so that the 
spin susceptibility is given by 
\be
\hat{{\chi}}_{\rm RPA}(\q, i \omega_n) = (\hat{{\bf 1}} - \hat{U}\hat{{\chi}}^0(\q,i \omega_n))^{-1} \hat{{\chi}}^0(\q,i \omega_n).
\ee
Here, $\hat{{\bf 1}}$ is a $2n^2 \times 2n^2$ identity matrix and the elements of block diagonal matrix 
$\hat{U}$ are  
\begin{eqnarray}
&&{U}_{\mu_1 \mu_2;\mu_3 \mu_4}  \nonumber\\
&&= \left\{
\begin{array}{@{\,} l @{\,} c}
U \,\,& (\mu_1=\mu_2=\mu_3=\mu_4)\\
U-2J \,\, & (\mu_1=\mu_2\ne\mu_3=\mu_4)\\
J \,\,& (\mu_1 = \mu_2\ne \mu_3 = \mu_4)\\
J \,\, & (\mu_1=\mu_4\ne \mu_2=\mu_3)\\
0 \,\,& (\mathrm{otherwise})
\end{array}\right.,  \nonumber\\
\end{eqnarray}
where $U^{\prime}$ = $U$ - $2J$ has been used as required by the rotational invariance. 
Analytic continuation $i \omega_n \rightarrow \omega + i \eta$ in all the calculations 
described below is performed with $\eta$ as 0.002eV. The unit of energy eV is used throughout 
unless stated otherwise.
\begin{figure}
\begin{center}
\vspace*{-2mm}
\psfig{figure=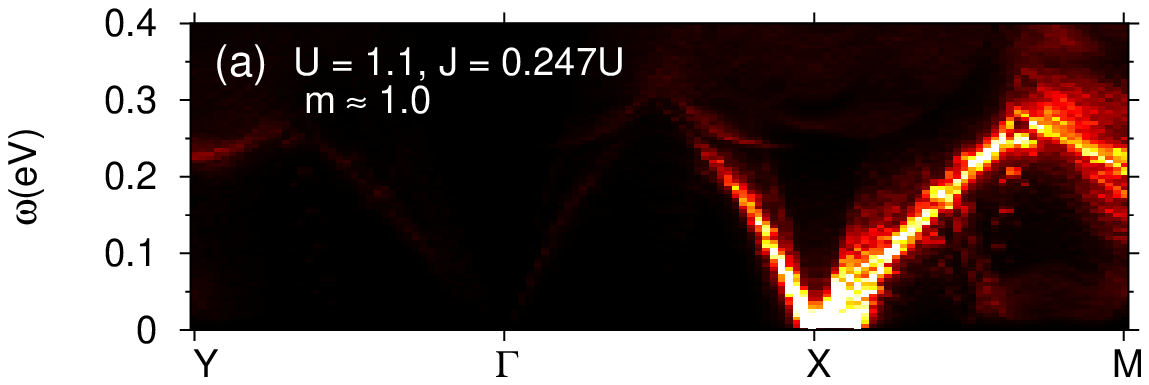,width=88mm,angle=0}
\vspace*{-0mm}
\psfig{figure=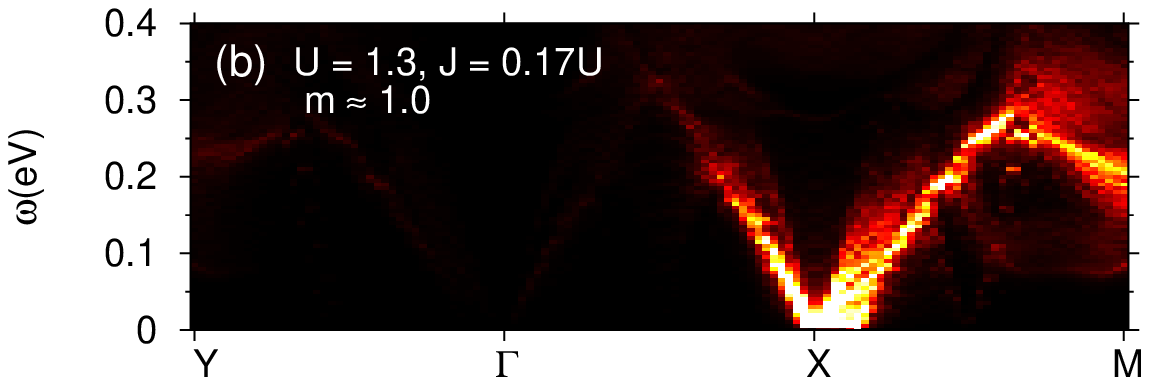,width=88mm,angle=0}
\vspace*{-0mm}
\psfig{figure=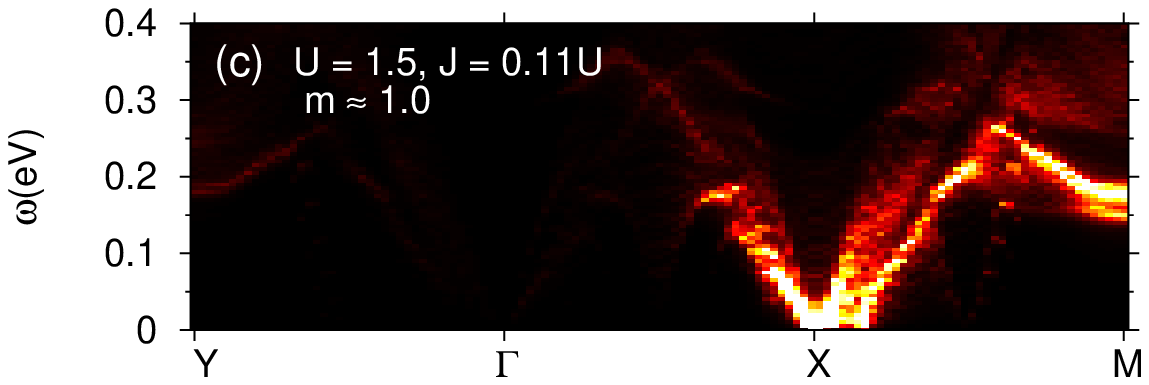,width=88mm,angle=0}
\vspace*{5mm}
\psfig{figure=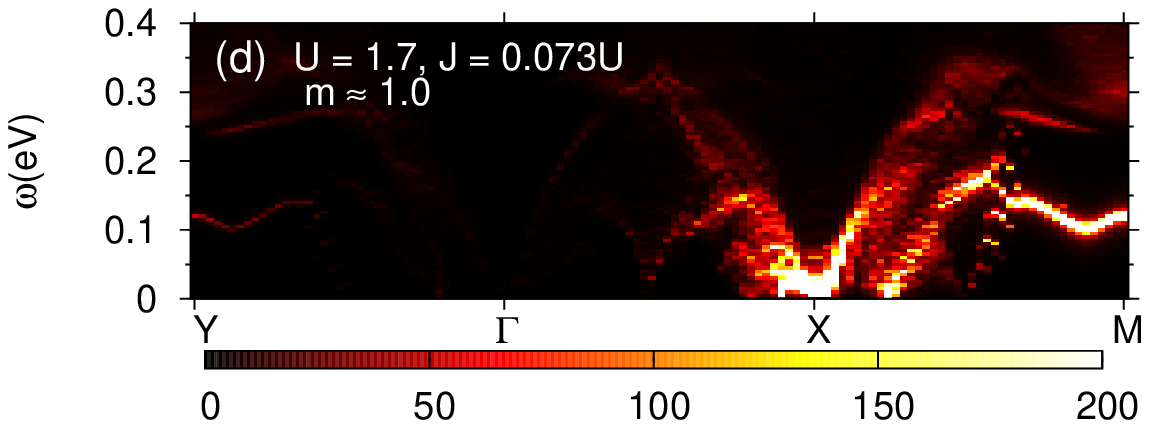,width=88mm,angle=0}
\vspace*{-13mm}
\end{center}
\caption{Imaginary part of the RPA-level physical spin susceptibility Im$\chi^{ps}_{\rm RPA}(\q, \omega)$
along the high-symmetry directions for the set of interaction parameters chosen 
in such a way that $m = 1$ in each case. (a) $U = 1.1, J = 0.247U$ (b) $U = 1.3, J = 0.167U$ (c) $U = 1.5, J = 0.110U$, and 
$U = 1.7,$ and $J = 0.730U$.}
\label{rpadisp}
\end{figure}  
\begin{figure}
\begin{center}
\vspace*{-2mm}
\hspace*{0mm}
\psfig{figure=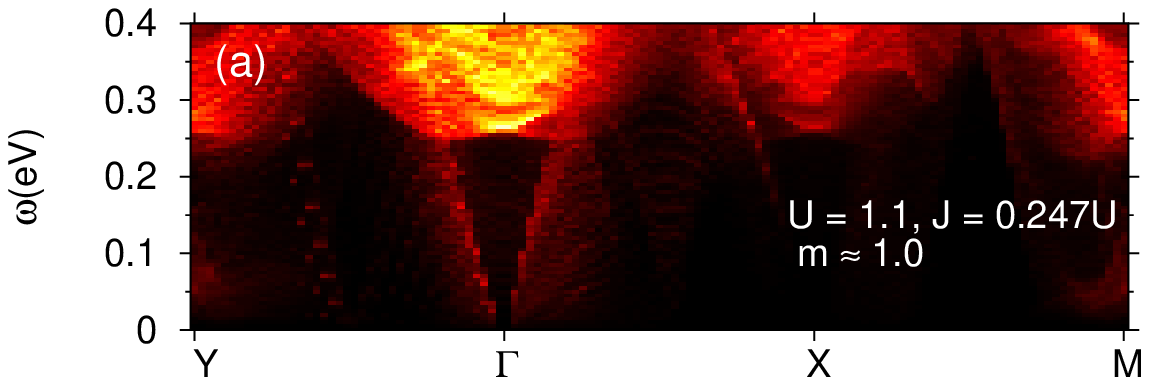,width=88mm,angle=0}
\psfig{figure=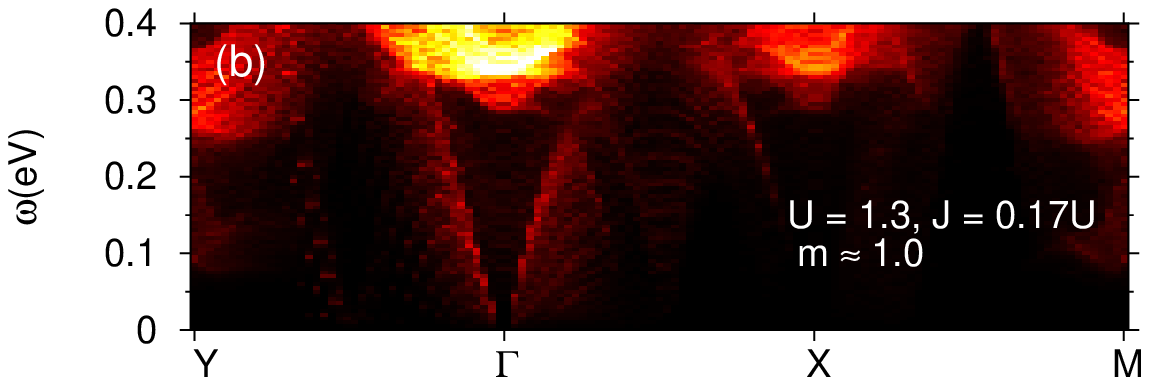,width=88mm,angle=0}
\psfig{figure=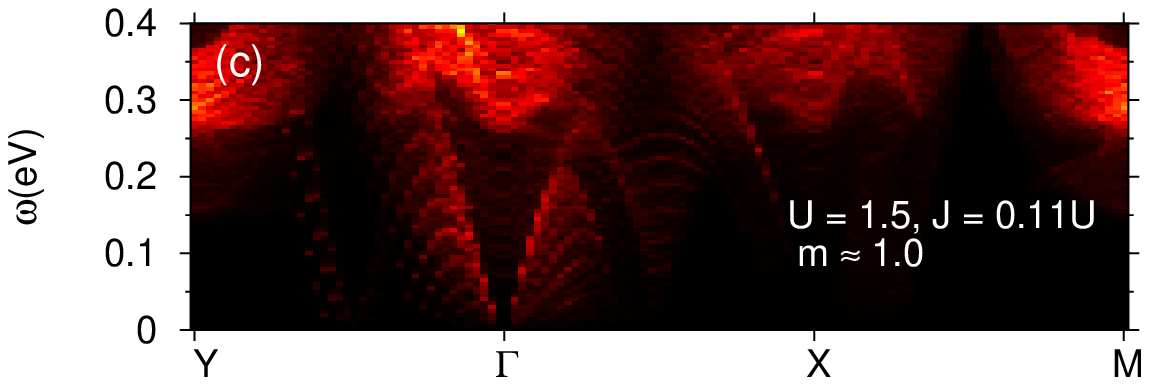,width=88mm,angle=0}
\vspace*{5mm}
\psfig{figure=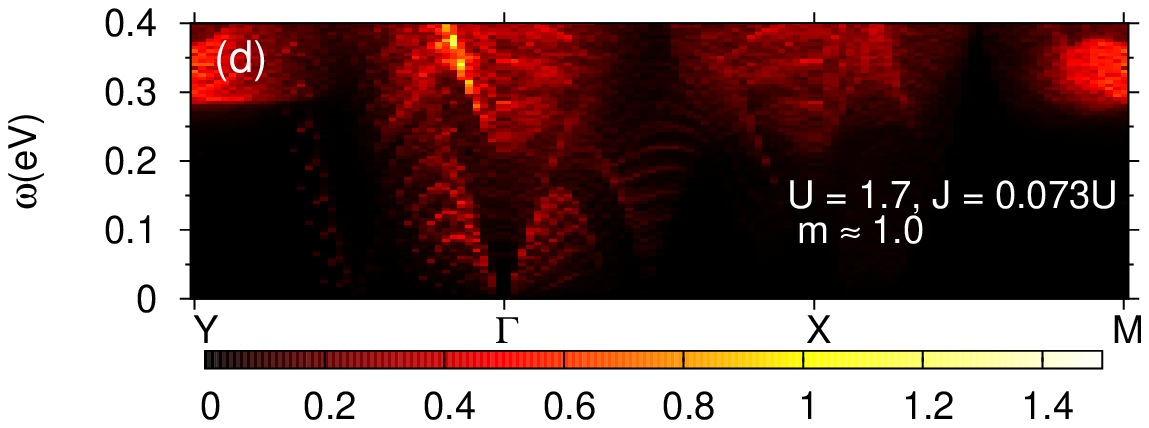,width=88mm,angle=0}
\vspace*{-13mm}
\end{center}
\caption{Imaginary part of the bare physical spin susceptibility Im$\chi^{ps}_{}(\q, \omega)$ along 
the high-symmetry directions for the set of interaction parameters
as in Fig. \ref{rpadisp}.}
\label{baredisp}
\end{figure}  
\section{Results}
\begin{figure}
\begin{center}
\vspace*{-4mm}
\hspace*{0mm}
\psfig{figure=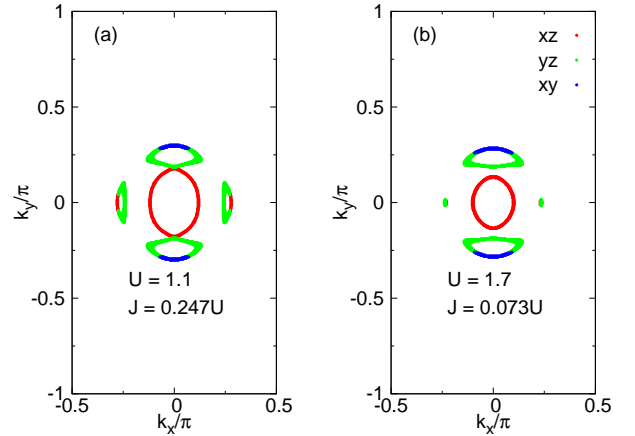,width=66.0mm,angle=-90}
\vspace*{-15mm}
\end{center}
\caption{FSs obtained for the two sets of interaction parameters (a) $U = 1.1, J = 0.247U$ and (b)
$U = 1.7, J = 0.073U$. Magnetic moment $m = 1$ in each case. In both the cases, FSs
include several pockets near $\Gamma$ with a difference in the sizes.}
\label{fs}
\end{figure}  
\begin{figure}
\begin{center}
\vspace*{-4mm}
\hspace*{-4mm}
\psfig{figure=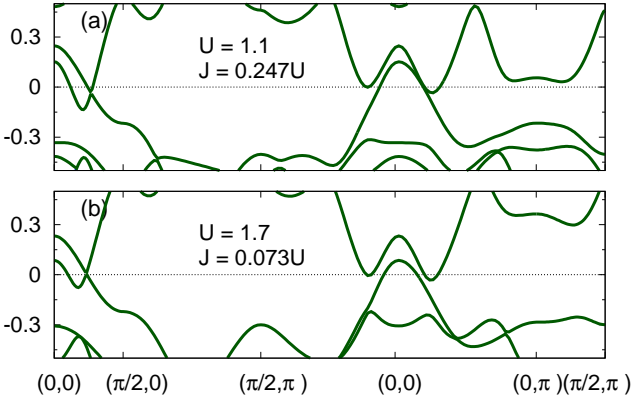,width=90.0mm,angle=0}
\vspace*{-10mm}
\end{center}
\caption{Electron dispersions for (a) $U = 1.1, J = 0.247U$ and (b)
$U = 1.7, J = 0.073U$ in the high-symmetry directions. Magnetic moment $m = 1$ in each case.}
\label{disp}
\end{figure}  
\begin{figure}
\begin{center}
\vspace*{-4mm}
\hspace*{0mm}
\psfig{figure=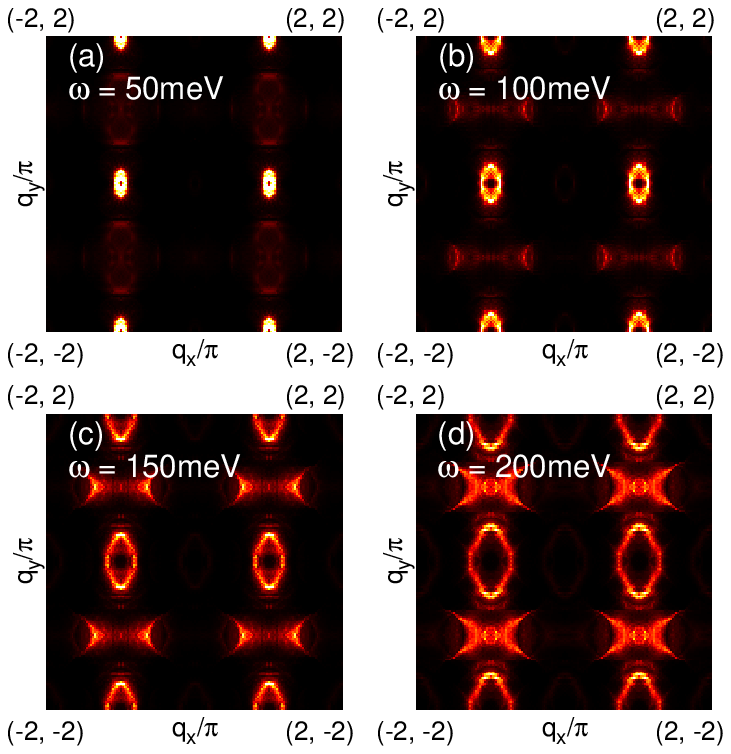,width=90.0mm,angle=0}
\vspace*{-12mm}
\end{center}
\caption{Constant energy cuts of Im$\chi^{ps}_{\rm RPA}(\q, \omega)$ for $U = 1.1, J = 0.247U$ at
energies (a) 50 meV, (b) 100 meV, (c) 150 meV, and (d) 200 meV.}
\label{cut1}
\end{figure}  
\begin{figure}
\begin{center}
\vspace*{-4mm}
\hspace*{0mm}
\psfig{figure=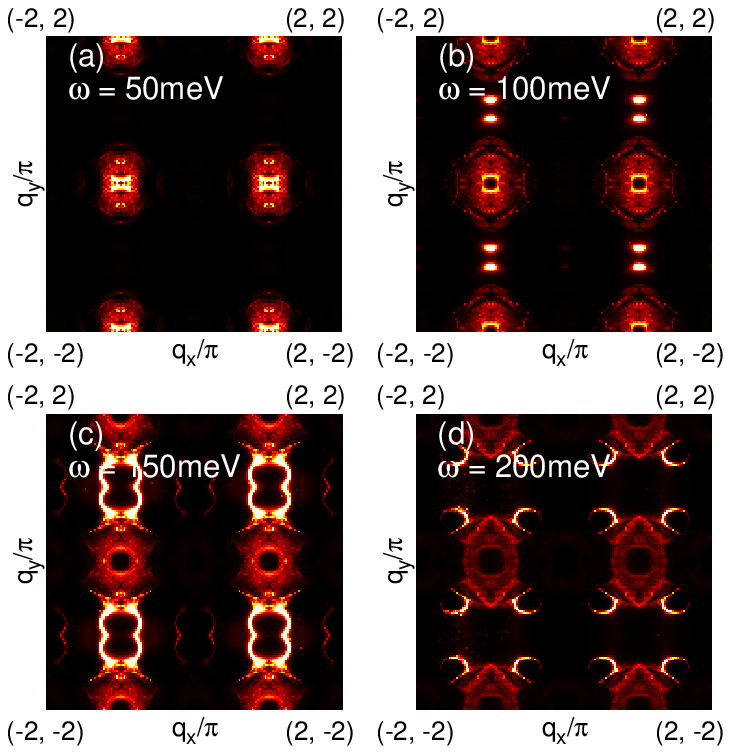,width=90.0mm,angle=0}
\vspace*{-12mm}
\end{center}
\caption{Constant energy cuts of Im$\chi^{ps}_{\rm RPA}(\q, \omega)$ for $U = 1.7, J = 0.073U$ at
energies (a) 50 meV, (b) 100 meV, (c) 150 meV, and (d) 200 meV.}
\label{cut2}
\end{figure}  
\begin{figure*}
\begin{center}
\vspace*{-4mm}
\hspace*{0mm}
\psfig{figure=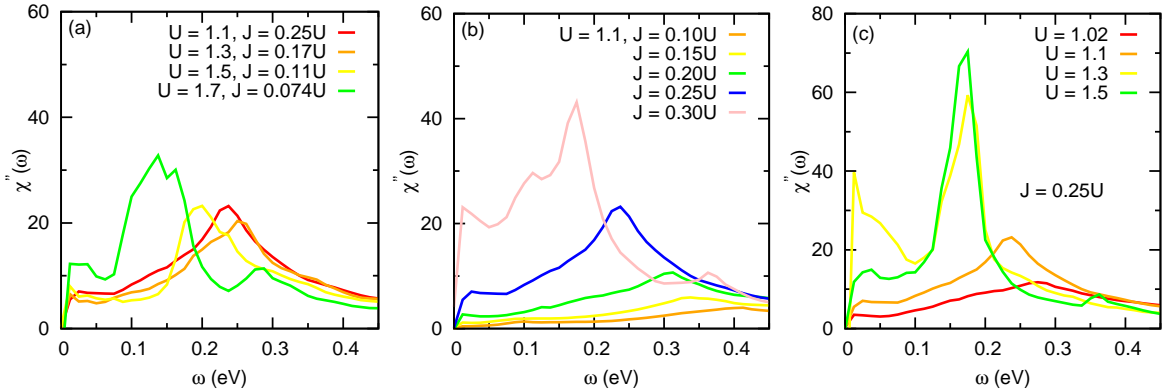,width=160mm,angle=0}
\vspace*{-5mm}
\end{center}
\caption{Spin-wave spectral functions for the cases when (a) magnetic moment $m = 1$ whereas $U$ and $J$ are set 
constant in (b) and (c), respectively.}
\label{spect}
\end{figure*}  
\begin{table}[ht]
\caption{Orbital resolved magnetic moments and charge densities for the two cases $U = 1.1$, $J = 0.247U$ 
and $U = 1.7$, $J = 0.073U$ denoted by subscripts 1 and 2, respectively.} 
\centering 
\setlength{\tabcolsep}{3pt} 
    \renewcommand{\arraystretch}{1} 
\begin{tabular}{c|c c c c c c c c c c c c c c c c} 
\hline\hline 
$ $ & $d_{3x^2-r^2}$ & $d_{xz}$ & $d_{yz}$ & $d_{xy}$ & $d_{x^2-y^2}$ \\ [0.5ex] 
$n_1$ &$1.457$ &  $1.230$ &  $1.165$ & $0.982$ &  $1.165$  \\
$n_2$ &$1.574$ &  $1.309$ & $1.189$ &  $0.834$ &  $1.094$  \\
$m_1$ &$0.160$ &  $0.155$ &  $0.252$ &  $0.338$ &  $0.098$  \\
$m_2$ &$0.099$ &  $0.185$ &  $0.197$ &  $0.422$ &  $0.075$  \\
[1ex] 
\hline 
\end{tabular}
\vspace{-2mm}
\end{table} 
\begin{table}[ht]
\caption{ $\sum_{\q,\omega}$Im $\chi^{sp}_{\alpha \alpha, \beta \beta}$ ($\q, \omega$) at RPA-level for
$U = 1.1$, $J = 0.247U$.} 
\centering 
\setlength{\tabcolsep}{3pt} 
    \renewcommand{\arraystretch}{1} 
\begin{tabular}{c|c c c c c c c c c c c c c c c c} 
\hline\hline 
$\alpha/\beta $ & $d_{3x^2-r^2}$ & $d_{xz}$ & $d_{yz}$ & $d_{xy}$ & $d_{x^2-y^2}$ \\ [0.5ex] 
$d_{3x^2-r^2}$ &$0.030$ &  $0.023$ &  $0.030$ & $0.039$ &  $0.019$  \\
$d_{xz}$ &  $0.023$     &  $0.076$ & $0.033$ &  $0.028$ &  $0.026$  \\
$d_{yz}$ &   $0.030$    &  $0.033$        &  $0.101$ &  $0.032$ &  $0.032$  \\
$d_{xy}$ &   $0.039$   &    $0.028$      &   $0.032$       &  $0.170$ &  $0.036$  \\
$d_{x^2-y^2}$ & $0.019$   &  $0.026$       &    $0.032$       &   $0.036$       &  $0.026$  \\
[1ex] 
\hline 
\end{tabular}
\vspace{-2mm}
\end{table} 

\begin{table}[ht]
\caption{Same as in Table II but for $U = 1.7$, $J = 0.073U$.} 
\centering 
\setlength{\tabcolsep}{3pt} 
    \renewcommand{\arraystretch}{1} 
\begin{tabular}{c|c c c c c c c c c c c c c c c c} 
\hline\hline 
$\alpha/\beta $ & $d_{3x^2-r^2}$ & $d_{xz}$ & $d_{yz}$ & $d_{xy}$ & $d_{x^2-y^2}$ \\ [0.5ex] 
$d_{3x^2-r^2}$ &$0.010$ &  $0.011$ &  $0.031$ & $0.017$ &  $0.009$  \\
$d_{xz}$ &  $0.011$      &  $0.085$ & $0.021$ &  $0.003$ &  $0.019$  \\
$d_{yz}$ &  $0.031$       &  $0.021$        &  $0.240$ &  $0.000$ &  $0.050$  \\
$d_{xy}$ &    $0.017$    &  $0.003$        &   $0.000$       &  $0.250$ &  $0.027$  \\
$d_{x^2-y^2}$ &  $0.009$  &   $0.019$      &      $0.050$     &   $0.027$       &  $0.040$  \\
[1ex] 
\hline 
\end{tabular}
\vspace{-2mm}
\end{table} 
Fig. \ref{rpadisp} shows the imaginary part of RPA susceptibility Im$\chi^{ps}_{\rm RPA}(\q, \omega)$ along the
high-symmetry directions Y-$\Gamma$-X-M calculated in SDW state for values of 
$U$ starting from 1.1 and increasing in step of 0.2 while $J$ is chosen
so that magnetic moment $m = 1$ with a maximum deviation only upto $\approx$ 2$\%$. We find a 
strong dependence of several features of the self-consistent mean-field state on $J$ as seen from the Table I, where the difference in
the orbital-resolved magnetizations for the two set of parameters $U = 1.1$, $J = 0.247U$ and $U = 1.7$, $J = 0.073U$ 
can be as large as $\approx$ 50$\%$ (for $d_{3x^2-r^2}$ orbital) despite the same magnetic moment. A significant 
difference in the two cases can be noticed for all the orbitals except $d_{x^2-y^2}$. Further, orbital-resolved 
magnetization decreases for $d_{xz}$ and $d_{xy}$ and increases for the remaining orbitals, when $J$ decreases.

Well-defined spin-wave excitations are obtained along $\Gamma$-X and in a part of X-M, where they can extend
up to $\approx$ 0.2eV for $U = 1.1$, $J = 0.247U$ as shown in the Fig. \ref{rpadisp}(a). Contributions to 
the excitations are mainly from the intraorbital susceptibilities particularly corresponding to 
the orbitals $d_{yz}$, $d_{yz}$ and $d_{xy}$ with latter contributing the most as 
seen from the Table II and Table III. Tables 
show integrated spectral weight $\sum_{\q,\omega}$Im$\chi^{ps}_{\rm RPA}(\q, \omega)$ with the
upper cutoff for the summation over $\omega$ chosen as $\omega_u = 0.5$eV. When $J$ is decreased, contributions
due to the interorbital susceptibilities also decrease whereas the contribution from the intraorbital 
susceptibility corresponding to $d_{yz}$ orbital increases and becomes similar in magnitude to that 
corresponding to $d_{xy}$.

The excitations become increasingly broad and
diffusive when $J$ is small. However, exactly opposite happens 
near M, where they become rather sharp and non-diffusive. At the same time, energy of
the spin-wave excitations near M increases. Heavy damping is present due to the metallicity 
of the SDW state as also reflected in the imaginary part of bare spin susceptibilities which appear gapless 
near $\Gamma$ but are gapped in other regions (Fig. \ref{baredisp}). The gap decreases in most part of 
the high-symmetry directions on decreasing $J$ except near 
M, where it increases on the contrary. Ungapped imaginary part of the bare susceptibility derives 
from the fact that FSs exist for all the sets of parameters considered with the Fermi pockets being 
clustered around $\Gamma$ (Fig. \ref{fs}). 

Behavior of the spin-excitations near M for smaller $J$ indicate sharpness may not be always plausible particularly in 
the multiorbital systems like pnictides, where many bands are located near the Fermi level. 
In order to understand the broadening especially near M when $J$ is large, it is useful to examine the reconstructed 
bands in the SDW state, wherefrom it can be seen 
that a part of the lowest lying partially-filled band gets lowered further near ($0, \pi$) on increasing $J$. 
Finally, it is in the immediate vicinity of the Fermi level for $U = 1.7$, $J = 0.073U$ as shown in Fig. \ref{disp}
and may affect the imaginary part of the bare and RPA-level susceptibilities significantly, which
are ungapped and heavily damped, respectively. At the same
time, other portions of the bands remain largely unaffected. Thus, the unusual feature of spin-wave excitations near M may 
arise due to a subtle interplay between the roles of interaction and the bandstructure. 

Fig. 5 and 6 show the energy cuts for $U = 1.1, J = 0.247U$ and 
$U = 1.7, J = 0.073U$, respectively. For the former case, anisotropy in the form of 
elliptical structure of excitations around X can be seen upto a energy as high as 200meV, where the major axis is oriented 
along $\Gamma$-Y. Excitations around M are rather broad and less intense with nature of 
anisotropy being similar to that around X. That changes quickly near 100meV when the structure around M becomes extended 
along $\Gamma$-X. In contrast, anisotropy around 
X is relatively weak for the latter set of interaction parameters and is strong only in a narrow energy window 
around $\omega = 150$meV for M. Moreover, the elliptical structure around X is also absent. Thus, anisotropy
near $\Q$ particularly in the elliptical form is very sensitive to $J$. In both cases, the 
anisotropy continues to exist upto high energy. As revealed also by the 
INS measurements, a significant anisotropy with elliptical shape is present
in the spin-wave excitations around X.\cite{diallo,zhao} 

Fig. \ref{spect} shows the spin-wave spectral functions calculated for three different cases. When the magnetic
moment $m \approx 1$ and a suitable set of interaction parameters is chosen, there is noticeable shift in the
spectral weight towards low-energy region upon increasing $U$ (Fig. \ref{spect} (a)). The hump-like peak 
structure located near 250meV for $U = 1.1, J = 0.247U$, which
is in agreement with the experiments, \cite{zhao1} relocates near 125meV $U = 1.7, J = 0.073U$. Another important factor that 
can be responsible for the spectral weight shifting towards low-energy region is doping of holes or electrons, which will 
be discussed elsewhere. On the other hand, transfer of the spectral weight towards
lower energy is significant in other two cases where one of the two interaction parameters is increased while keeping the other constant.
\section{Conclusions}
To conclude, we have investigated the role of Hund's coupling in the spin-wave excitations of the SDW 
state of undoped iron pnictides by using a realistic electronic structure within a five-orbital model. We find that the Hund's coupling at the higher end of the range of various theoretical and experimental estimates ($J \sim U/4$) is
required for the sharp and well-defined spin-wave dispersion in most part of the
high-symmetry directions for a given magnetization. Not only that a similar value of Hund's coupling
is also crucial for the elliptical
shape of the anisotropy around $\Q$ = ($\pi$, 0) in the spin-wave excitations as well as 
for the spectral weight to be concentrated near energy $\gtrsim$ 200meV. Thus, our study highlights the essential role of Hund’s coupling
in describing the experimentally observed features of spin-wave excitations in the SDW state of undoped iron pnictides.
\section{Acknowledgement}
We acknowledge the use of HPC clusters at HRI.

\end{document}